\documentclass{article}
\usepackage[margin=3cm]{geometry}

\usepackage{authblk}

\usepackage{amsmath,amssymb,amsfonts}
\usepackage{mathtools} 
\usepackage{physics}
\usepackage{makecell}
\setcellgapes{1ex}
\usepackage{cases}
\usepackage{booktabs} 
\usepackage{multirow}

\usepackage{graphicx,color}
\usepackage{epstopdf}
\usepackage{caption}
\usepackage[nocompress]{cite}

\usepackage[shortlabels]{enumitem}

\usepackage{caption}
\usepackage{subcaption}

\usepackage{float}

\usepackage{todonotes}

\numberwithin{equation}{section}

\usepackage{hyperref}
\hypersetup{colorlinks=true,allcolors=blue,urlcolor=cyan}

\title{The choice of variables in cosmological dynamical systems}

\author[1,2]{Christian G. B\"ohmer\footnote{Email: c.boehmer@ucl.ac.uk}}
\author[1]{Antonio d'Alfonso del Sordo\footnote{Email: a.dalfonsodelsordo@ucl.ac.uk}}

\affil[1]{Department of Mathematics, University College London, \authorcr Gower Street, London WC1E 6BT, UK\medskip}
\affil[2]{Astrophysics Research Centre, School of Mathematics, \authorcr Statistics and Computer Science, University of KwaZulu-Natal, \authorcr Private Bag X54001, Durban 4000, South Africa\medskip}

\date{12 May 2026} 

\pagecolor{white}

\begin{document}

\renewcommand{\arraystretch}{1.2} 
\setlength{\tabcolsep}{1ex} 
\setlength{\extrarowheight}{1ex} 

\maketitle

\begin{abstract}
Dynamical systems techniques are a powerful tool to analyse systems of ordinary differential equations, written in an appropriate form. For a given theory of gravity, the cosmological field equations typically lead to a system of ordinary differential equations. Casting these cosmological equations into the form of a dynamical system requires a careful choice of the dynamical variables. Despite this being a critical step, relatively little is said about this process in the literature. We discuss how different variable choices affect the information that can be extracted from the Friedmann equations. We begin by reviewing the standard cosmological model with dark matter, radiation, and dark energy, and include quintessence models. We revisit well-known models with an exponential potential using new variables. This discussion is then extended to models with scalar fields and more intricate coupling terms.
\end{abstract}

\tableofcontents

\section{Introduction}

Modern cosmology aims to understand the origin and evolution of the Universe by reconciling the predictions of the Einstein field equations with observational data. The Einstein field equations are a system of ten coupled, nonlinear partial differential equations, which are the equations of motion of gravity. Under the assumptions of homogeneity and isotropy of the universe (collectively known as the cosmological principle), the system is reduced to a finite-dimensional system of nonlinear ordinary differential equations (ODEs).

Any system of ODEs can be cast into the form of a dynamical system; by this, we mean a first-order system of equations of the form
\begin{align}
    \label{dyn1}
    \dot{\boldsymbol{x}} := \frac{\mathrm{d}\boldsymbol{x}}{\mathrm{d}t} = 
    \boldsymbol{f}(\boldsymbol{x},t) \,,
\end{align}
where $t\in\mathbb{R}$ is the independent variable (cosmological time in the sections that follow), and $\boldsymbol{x}=(x_1,x_2,\ldots,x_n) \in X$ represents the state of the system within the phase space $X \subseteq \mathbb{R}^n$. If the function $\boldsymbol{f}$ does not depend on time, the system is called autonomous. Non-autonomous systems can always be made autonomous by defining an additional state variable. To ensure that a unique solution exists for any initial condition, we require $\boldsymbol{f}:X\to \mathbb{R}^n$ to be at least Lipschitz continuous (though in most physical applications, $\boldsymbol{f}$ is assumed to be continuously differentiable). We view the function $\boldsymbol{f}$ as a vector field in $\mathbb{R}^n$, with components $\boldsymbol{f}(\boldsymbol{x}) =(f_1(\boldsymbol{x}), f_2(\boldsymbol{x}), \ldots, f_n(\boldsymbol{x}))$. The vector field $\boldsymbol{f}$ generates a flow $\boldsymbol{\varphi}_t:X\to X$ such that $\boldsymbol{x}(t)=\boldsymbol{\varphi}_t(\boldsymbol{x}_0)$ represents the trajectory passing through the initial point $\boldsymbol{x}_0\in X$. 

Points $\boldsymbol{x}^{*}$ that satisfy $\boldsymbol{f}(\boldsymbol{x}^{*})=\boldsymbol{0}$ are known as fixed points or critical points. To understand the dynamics of the system near a fixed point, we linearise the system around that point. The dynamics of the linearised system are qualitatively equivalent to the original system, and information about the stability of a critical point is contained in the eigenvalues of the matrix $\boldsymbol{\nabla}\boldsymbol{f}(\boldsymbol{x}^{*})$, known as the Jacobian matrix or stability matrix. If all eigenvalues of the Jacobian matrix have positive real parts, the point is unstable or a repeller. If all eigenvalues have negative real parts, the point is stable and called an attractor. Lastly, if the eigenvalues have real parts with opposite signs, then the corresponding fixed point is called a saddle point. If at least one eigenvalue is zero, the point is called non-hyperbolic and linear stability theory cannot be applied. A broad body of applied mathematics literature explores the properties of dynamical systems~\cite{Wiggins2003,Strogatz2024}. 

Over the past two decades, dynamical systems methods have become standard tools to study cosmological equations~\cite{wainwright1997,coley2003,leon2011}; for a review focusing on dynamical dark energy models, see~\cite{bahamonde2018}. Linear stability analysis is the main tool employed in these works; however, more advanced techniques, such as the Lyapunov function method, centre-manifold techniques, and global phase–space analysis, have also been considered. 

The key insight is simple yet powerful: the cosmological evolution is represented by trajectories in a suitably defined state space. The fixed points of the autonomous system of ODEs correspond to cosmological epochs such as inflation, radiation domination, matter domination, and late-time acceleration. The cosmological evolution
\begin{align*}
    \text{inflation} \longrightarrow 
    \text{radiation} \longrightarrow
    \text{matter} \longrightarrow
    \text{dark energy}
\end{align*}
can now be interpreted as a sequence of heteroclinic orbits, i.e.\ trajectories which join a pair of distinct fixed points. Although not all theoretical models will contain these specific points, those that fail to reproduce this qualitative sequence are generally of little interest to physical cosmology. Consequently, candidate models can be constrained by analysing their phase space structure: viable cosmologies generally require the presence of fixed points associated with radiation, matter, and accelerated eras, together with suitable connecting trajectories. However, since fixed points correspond to asymptotic behaviour, the physical evolution is more accurately described by transient trajectories in their vicinity. 

\section{Standard cosmology with matter, radiation and dark energy}
\label{sec:standarcosmology}

As a first example, we discuss the cosmological field equations in the presence of matter, radiation, and a cosmological constant. Motivated by observational evidence~\cite{PlanckCollabVI, efstathiou2020}, we assume a spatially flat Friedmann--Lema\^itre--Robertson--Walker (FLRW) metric, which will be adopted throughout this work. This simple model approximately describes our Universe and captures its main qualitative features. Further details may be found in \cite{Boehmer2017a,Boehmer2017}. The spatially flat cosmological Einstein field equations are given by
  \begin{align}
    3H^2 - \Lambda &= 
    \kappa\, \sum_i \rho_i \,,
    \label{field1a}\\
    -2\dot{H} - 3H^2 + \Lambda &= 
    \kappa\, \sum_i p_i \,,
    \label{field2a}
  \end{align}
where $H(t)=\dot{a}/a$ is the Hubble function, $a(t)$ is the scale factor, $\Lambda$ is the cosmological constant, $\kappa=8\pi G/c^4$ is a coupling constant which determines the strength of the gravitational force and connects the Newton's gravitational constant $G$ and the speed of light $c$, $\rho_i$ and $p_i$ are the energy density and pressure of some matter components, labelled by the index $i$. We assume that all matter sources satisfy their respective conservation equation
\begin{align}
  \dot{\rho}_i + 3\frac{\dot{a}}{a}(\rho_i + p_i) = 0 \,.
  \label{cons1}
\end{align}
We consider these sources to be a perfect fluid with a constant equation-of-state parameter $w_i$, which relates the pressure and the energy density via the equation of state $p_i = w_i \rho_i$. For example, for dust or pressureless matter, one has $w=0$; for radiation, one has $w=1/3$.

\begin{figure}[!htb]
    \centering
    \includegraphics[width=0.65\textwidth]{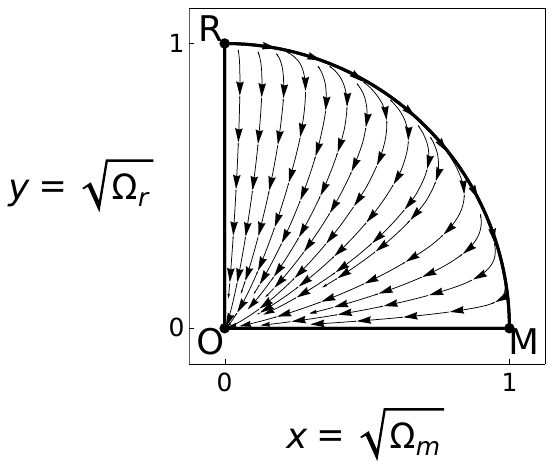}
    \caption{Phase portrait of the dynamical system in Eq.~(\ref{eq4b}). The fixed points are Point $R$, Point $M$, and Point $O$.}
    \label{fig:2dstandard}
\end{figure}

It is convenient to introduce the dimensionless density parameters
\begin{align}
  \Omega_i = \frac{\kappa\, \rho_i}{3H^2}, \qquad
  \Omega_{\Lambda} = \frac{\Lambda}{3H^2}\,,
  \label{density}
\end{align}
for the different sources of matter. We note that these variables are well defined only for  $H \neq 0$, and thus exclude the Einstein static universe or bouncing cosmological models; in principle, one could work with other variables, see for example~\cite[Sec.~3.4]{bahamonde2018}, which allow for non-spatially flat models where $H=0$ naturally appears. Let us denote the respective energy densities of matter and radiation by $\Omega_{\rm m}$ and $\Omega_{\rm r}$, and assume that there are no other matter sources. These quantities are observable, see for example~\cite{PlanckCollabVI}, and measure the relative energy content of these matter components. Then, Eq.~(\ref{field1a}) becomes the algebraic equation
\begin{align}
  1 = \Omega_{\rm m} + \Omega_{\rm r} + \Omega_{\Lambda} \,,
  \label{ex:c1}
\end{align}
which is often called the \emph{constraint equation}. It shows that only two of the three densities are independent. The field equations, Eqs.~(\ref{field1a})~and~(\ref{field2a}), together with the conservation equation, Eq.~\eqref{ex:c1}, can be re-written as a system of two first-order ODEs, which describes the dynamics of the universe; for example, see~\cite{Boehmer2017, bahamonde2018}. As our primary interest lies in the choice of variables, we employ a slightly different, more natural approach. Since the density parameters for matter and radiation are non-negative physical quantities, i.e.\ $\Omega_{\rm m}>0$ and $\Omega_{\rm r}>0$, we introduce the variables $x^2 = \Omega_{\rm m}$ and $y^2 = \Omega_{\rm r}$. Assuming a positive cosmological constant $\Lambda > 0$, so that $\Omega_{\Lambda} > 0$, in the constraint equation, Eq.~(\ref{ex:c1}), yields
\begin{align}
  0 \leq x^2 + y^2 = 1 - \Omega_{\Lambda} \leq 1 \,.
  \label{ex:c2}
\end{align}
This means that the $(x,y)$ phase space coincides with the sector of the unit circle in the first quadrant. In these variables, by using the approach outlined in~\cite{Boehmer2017}, the dynamical equations read
\begin{align}
  x' = \frac{1}{2} x (3 x^2+4 y^2-3) \,, \qquad
  y' = \frac{1}{2} y (3 x^2+4 y^2-4) \,,
  \label{eq4b}
\end{align}
where the prime denotes differentiation with respect to $N=\log a$. The fixed points are easily found to be Point $O=(0,0)$, Point $R=(0,1)$, and Point $M=(1,0)$. These correspond to a dark-energy-dominated universe, a radiation-dominated universe, and a matter-dominated universe, respectively. The phase space diagram of system in Eq.~(\ref{eq4b}) is shown in Fig.~\ref{fig:2dstandard}. Linear stability analysis shows that Point $R$ is unstable and can be interpreted as the early-time repeller, meaning that all trajectories emerge from this point. On the other hand, Point $O$ is the late-time attractor: all trajectories eventually approach this point. Lastly, Point $M$ is a saddle point: trajectories approach it along stable directions but are ultimately repelled along unstable directions, towards Point $O$.

It is interesting to note that Eqs.~(\ref{eq4b}) are of the form of a Kolmogorov-type population model, which means $\dot{x}=x A(x,y)$ and $\dot{y}=y B(x,y)$, where $A$ and $B$ are continuously differentiable functions. Kolmogorov models emerged as natural generalisations of Lotka--Volterra predator-prey model. In this setting, the competitive species are represented by the two fluid components of the universe and the dark energy (modelled by the cosmological constant); for more details on Lotka--Volterra systems in cosmology, see~\cite{Perez2014}. Through the language of population dynamics, one can think of the matter sources as species in a predator-prey model: the species associated with $\Omega_{\Lambda}$ is dominant at late times, while the species $\Omega_{\rm m}$ and $\Omega_{\rm r}$ eventually disappear.

The cosmological constant $\Lambda$ is a good model for the late-time accelerated expansion of the Universe. However, from a theoretical perspective, it poses well-known problems, particularly the large discrepancy between its observed value and theoretical estimates of the vacuum energy density. This motivates the introduction of a scalar field $\phi$ with potential energy $V(\phi)$ as a dynamical model of dark energy. Models of this type do not necessarily solve the cosmological constant problem but have several interesting properties, such as scaling solutions. For sufficiently steep potentials, the system evolves towards a steady state where the scalar field `tracks' matter or radiation. This shows that the cosmological evolution with scalar fields can be insensitive to initial conditions and allows scalar degrees of freedom to remain dynamically relevant over long epochs without fine-tuning.
Models of this type are often called quintessence models~\cite{amendola2000,copeland2006}, and we will discuss these in the following sections. 

\section{Quintessence models -- old and new variables}\label{sec:copelandxsigma}

In 1998, Copeland, Liddle, and Wands~\cite{copeland1998} introduced the dimensionless variables 
\begin{equation}
   x=\frac{\kappa\dot{\phi}}{\sqrt{6} H}\,,\quad y=\frac{\kappa \sqrt{V}}{\sqrt{3}H}\,,\quad \Omega=\frac{\kappa^2\rho}{3H^2}\,,
    \label{eq:copelandvars}
\end{equation}
assuming that the potential is exponential, i.e.\ $V=V_0\exp(-\lambda \kappa\phi)$, for some constants $V_0$ and $\lambda$, in order to analyse quintessence models with a scalar-field potential. Here $\rho$ denotes the total background matter energy density, distinct from the scalar-field contributions. Let us remark that the dynamical variables are normalised by $H$, or $H^2$, and therefore only well defined when $H\neq 0$. Note that $H>0$ describes an expanding universe whereas $H<0$ corresponds to contracting solutions. Trajectories approaching $H=0$ correspond to divergences in the variables, so this formalism does not directly describe turning points of the scale factor (such as bounces or recollapses). By normalising the scalar-field's kinetic energy term and potential energy term with respect to the Hubble parameter, these variables allow us to express the cosmological field equations as a dynamical system. The Friedmann constraint takes the form
\begin{align}
  0 \leq x^2 + y^2 = 1 - \Omega \leq 1 \,,
  \label{ex:copeland2}
\end{align}
similar to Eq.~\eqref{ex:c2}. The main difference here is that the variable $x$ is no longer necessarily positive which means that the $(x,y)$ phase space becomes the half disc with $y \geq 0$.

The appeal of these variables lies in several key features. Among these, they compactify the physically relevant phase space, making it easier to identify global properties of the solutions and enabling a straightforward classification of the critical points, and hence interpret the dynamics of the universe.

In subsequent studies of modified gravity models with interaction terms between dark matter and dark energy involving geometric quantities (such as bulk and boundary term couplings at the level of the action or other terms at the level of the field equations), the cosmological equations acquire a significantly more complicated structure, yet the same set of variables are found to be effective; see for example~\cite{Boehmer:2024rqk,Ashi:2025dba}. For models of this type, the constraint equation takes the form
\begin{align}
  x^2 + y^2 + \Omega - x f(y,\Omega) = 1 \,.
  \label{ex:copeland3}
\end{align}
Here $f(y,\Omega)$ denotes a given function that specifies the model. In the simplest case, where $f$ is constant, a renormalised version of the standard variables introduced in~\cite{copeland1998} can be employed, and the resulting system retains a structure closely analogous to minimally coupled quintessence~\cite{Boehmer:2024rqk}. However, when $f(y,\Omega)$ is an arbitrary smooth function of its arguments, the constraint, given in Eq.~\eqref{ex:copeland3}, introduces significant difficulties in the analysis. To be able to work with concrete models, we consider the specific interaction function $f= ky^{-\alpha}\Omega^{\alpha/2}$ such that the constraint becomes
\begin{equation}
    x^2 + y^2 + \Omega - k x \frac{\Omega^{\alpha/2}}{y^\alpha} = 1 \,,
    \label{eq:intterm0}
\end{equation}
where $\alpha$ is a fixed power and $k$ is a coupling constant. The particular choice $\alpha = 2$ remains tractable using the standard $(x,y)$ variables; for a detailed analysis, see~\cite{Boehmer:2024rqk}. This follows because the constraint is linear in the matter variable $\Omega$. However, the case $\alpha = -2$ requires a different treatment. As we show later, this can be achieved by solving the Friedmann constraint for $y$ rather than for $\Omega$; an approach first discussed in~\cite{dAlfonso2025}.

To establish a baseline model against which our results can be compared, let us begin by considering the quintessence model with an exponential potential studied in~\cite{copeland1998}, but solve the Friedmann constraint for $y$ instead of $\Omega$. The resulting system captures the same dynamical behaviour, albeit from a slightly different perspective. As in the previous discussion, we work with $\sigma^2 = \Omega$ to ensure that the matter variable remains non-negative. Using the variables defined in Eq.~\eqref{eq:copelandvars}, the Friedmann constraint again takes the form $1=x^2+y^2+\sigma^2$. In the case of an exponential potential, we have that $V \geq 0$, which yields $y \geq 0$. Then, $1\geq 1-x^2-\sigma^2=y^2\geq 0$. Hence, the physical phase space is contained in the unit circle $0\leq x^2+\sigma^2\leq 1$. Furthermore, we are only interested in non-negative matter densities $\sigma\geq 0$, therefore we can restrict our analysis to the $(x,\sigma)$ semi circle with $\sigma \geq 0$. The dynamical equations are
\begin{align}
    x' &= \frac{3}{2} x \left(\sigma ^2 (w+1)+2 x^2-2\right) -
    \sqrt{\frac{3}{2}} \lambda \left(\sigma ^2+x^2-1\right), 
    \label{eq:copxsigma1}\\
    \sigma' &= \frac{3}{2} \sigma \left((\sigma ^2-1) (w+1)+2 x^2\right).
    \label{eq:copxsigma2}
\end{align}
The properties of the dynamical system depend on the values of the parameters $\lambda$ and $w$. Physical variables of interest, like the deceleration parameter $q$, can be expressed in terms of the dynamical variables
\begin{equation}
    q:=-1-\frac{\dot{H}}{H^2}=\frac{3}{2} (w+1) \sigma^2+3 x^2-1 \,.
    \label{eq:accelerationcopeland}
\end{equation}
Integrating Eq.~\eqref{eq:accelerationcopeland}, one can evaluate $a(t)$ at any fixed point $(x_0,\sigma_0)$ of the phase space, that is,
\begin{equation}
    a(t) = a_0 (t-t_0)^{2\big/\left(6x_0^2+3(1+w)\sigma_0^2\right)} \,, \label{eq:powerlawcopeland}
\end{equation}
where $a_0$ and $t_0$ are constants of integration. Note that we can define the effective (or total) energy density and pressure of the system as
\begin{align}
    \widetilde{\rho}=\rho+\frac{1}{2}\dot{\phi}^2+V \,, \qquad
    \widetilde{p}=p+\frac{1}{2}\dot{\phi}^2-V \,.
\end{align}
Using the acceleration equation, Eq.~\eqref{eq:accelerationcopeland}, the effective equation-of-state parameter $\widetilde{w}$ in terms of the dimensionless variables is
\begin{equation}
    \widetilde{w}=\frac{\widetilde{p}}{\widetilde{\rho}}=-1 + 2 x_0^2 + (1 + w) \sigma_0^2 \,.
    \label{eq:weffcopxsigma}
\end{equation}
This allows us to re-write Eq.~\eqref{eq:powerlawcopeland} in the convenient form
\begin{equation}
    a = a_0 \left(t-t_0\right)^{2\big/\left(3(1+\widetilde{w})\right)} \,.
    \label{eq:aeqnforcopelandxsigma}
\end{equation}
The fixed points of the dynamical system given in Eqs.~\eqref{eq:copxsigma1} and~\eqref{eq:copxsigma2} are listed in Table~\ref{tab:fixedpointscopelandxsigma}.

\begin{table}[!b]
\centering
{\begin{tabular}{@{}lccccc@{}}\hline
        Point & Coordinates $(x,\sigma)$ & Existence & $\widetilde{w}$ & $q$ \\\hline 
        Point $A_{-}$ & $(-1,0)$ & for all $\lambda$, $w$ & $1$ & $2$ 
        \\
        Point $A_{+}$ & $(1,0)$ & for all $\lambda$, $w$ & $1$ & $2$ \\
        Point $B$ & $\displaystyle\left(\sqrt{\frac{3}{2}}\frac{w+1}{\lambda },\frac{\sqrt{\lambda ^2-3 w-3}}{\lambda }\right)$ & $\lambda^2\geq 3(1+w)$ & $w$ & $(3 w+1)/2$\\
        Point $C$ & $\left(\lambda/\sqrt{6},0\right)$ & for all $\lambda,w$ & $-1+\lambda^2/3$ & $-1 + \lambda ^2/3$ \\
        Point $D$ & $\left(0,1\right)$ & for all $\lambda$, $w$ & $w$ & $(3 w+1)/2$
        \\\hline
    \end{tabular}}
\caption{Fixed points for Eqs.~\eqref{eq:copxsigma1} and~\eqref{eq:copxsigma2} and values of $\widetilde{w}$ and deceleration parameter $q$.}
\label{tab:fixedpointscopelandxsigma}
\end{table}

We observe the following features. At Points $A_{\pm}$, the universe is dominated by the kinetic energy of the scalar field (i.e.\ $x^2=1$); the effective equation of state corresponds to that of a stiff fluid (i.e.\ $\widetilde{w}=1$), and no acceleration occurs, as indicated by the deceleration parameter $q=2$. Moreover, from Eq.~\eqref{eq:aeqnforcopelandxsigma}, the scale factor evolves as $a\propto t^{1/3}$. The existence of these fixed points is guaranteed for all values of $\lambda$ and $w$. Point $B$ represents a scaling solution, for which the effective equation of state coincides with that of the matter component. For physically acceptable values of $w$, we have $q<0$, implying accelerated expansion only when $-1\leq w<-1/3$; otherwise, the expansion is decelerating. Point $C$ corresponds to a universe entirely dominated by the scalar field. In contrast to its analogue in the $(x,y)$ coordinates, this point lies within the unit circle only when $\lambda<\sqrt{6}$. The corresponding effective equation of state yields accelerated expansion for $\lambda^2<2$, corresponding to a power-law expansion regime with $\ddot{a}>0$. As $\lambda\rightarrow 0$, the solution approaches a de Sitter expansion dominated by a cosmological constant. Point $D$ behaves analogously to the origin in the $(x,y)$ coordinates and corresponds to a matter-dominated universe with $\sigma^2=1$. Its existence is guaranteed for all values of $\lambda$ and $w$. At this point the effective equation of state reduces to the matter equation-of-state parameter, and the cosmic expansion is therefore dominated by the matter component rather than the scalar field. The stability properties of the fixed points are summarised in Table~\ref{tab:eigenvaluescopelandxsigma}.

\begin{table}[!h]
\centering
{\begin{tabular}{@{}lcl@{}}\hline
        Point & Eigenvalues & Classification  \\\hline 
        Point $A_{-}$ & $-\frac{3}{2} (w-1)\,,\quad 6+\sqrt{6} \lambda$& 
        \makecell[l]{
             unstable if $\lambda\geq -\sqrt{6}$ \\
             saddle if $\lambda<-\sqrt{6}$
        }\\[2ex]  
        Point $A_{+}$ & $-\frac{3}{2} (w-1)\,,\quad 6-\sqrt{6} \lambda$& 
        \makecell[l]{
             unstable if $\lambda\leq \sqrt{6}$ \\
             saddle if $\lambda>\sqrt{6}$
        }\\[1ex]  
        Point $B$ & 
        $\frac{3}{4\lambda}\left[\lambda(w-1)\pm\delta\right]$ & 
        \makecell[l]{
             stable node if $3(1+w)<\lambda^2<\frac{24(w+1)^2}{9w+7}$ \\[1ex] 
             stable spiral if $\lambda^2\geq \frac{24(w+1)^2}{9w+7}$
        }\\[3ex]  
        Point $C$ & 
        $\frac{1}{2} \left(\lambda ^2-6\right),\quad \frac{1}{2} \left(\lambda ^2-3 w-3\right)$ & 
        \makecell[l]{
             stable if $\lambda^2<3(1+w)$ \\
             saddle if $3(1+w)\leq \lambda^2<6$
        } \\[2ex]
        Point $D$ &$\frac{3}{2}(w-1)\,,\quad 3(w+1)$& saddle
        \\\hline
    \end{tabular}}
\caption{Stability of the critical points for the system given in Eqs.~\eqref{eq:copxsigma1}~and~\eqref{eq:copxsigma2}.\\ Here $\delta=\sqrt{(w-1)\left[(9w+7)\lambda^2-24(w+1)^2\right]}$.}
\label{tab:eigenvaluescopelandxsigma}
\end{table}

\begin{figure}[!h]
    \centering
    \includegraphics[width=0.9\textwidth]{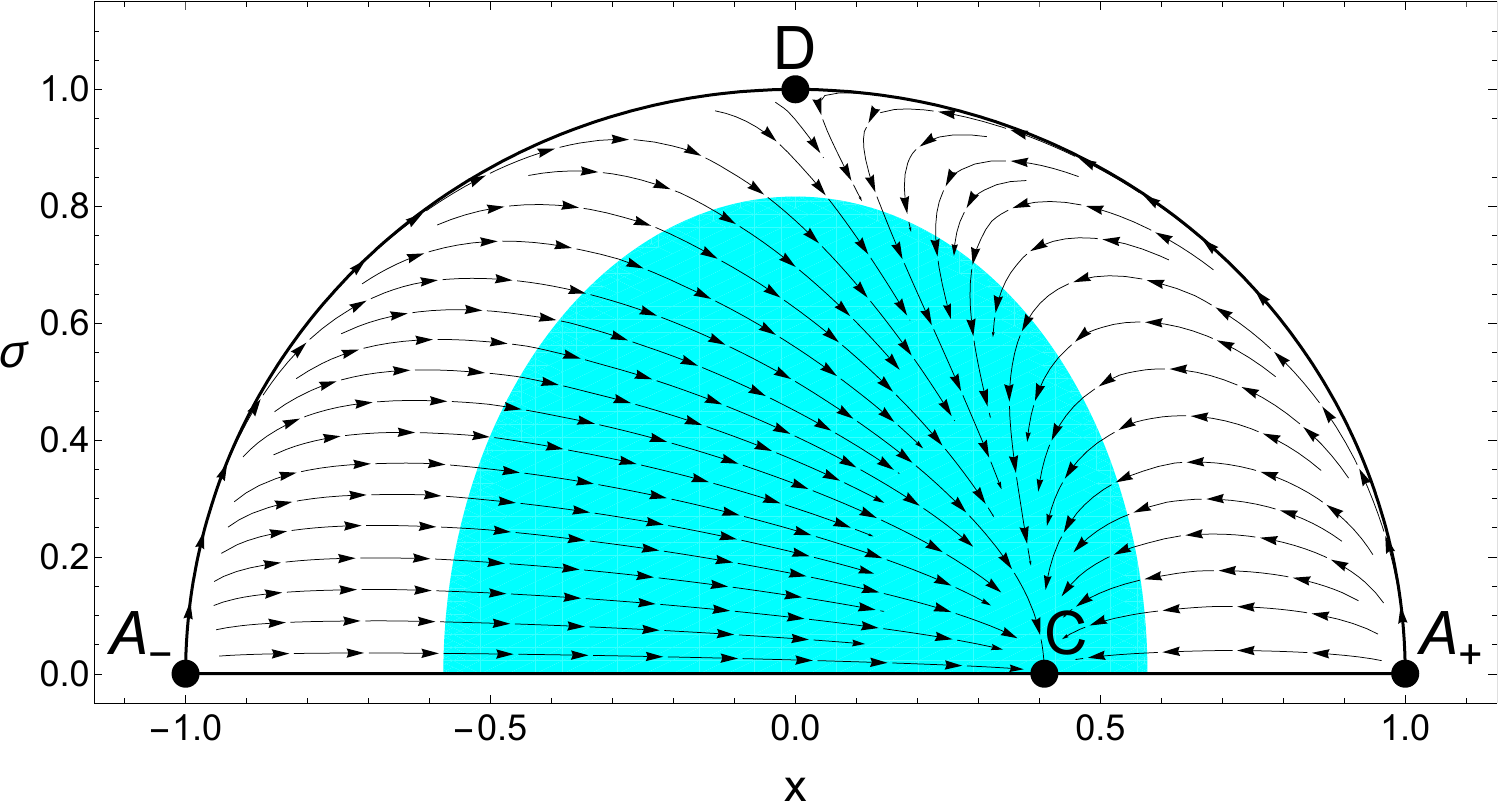}
    \caption[Phase space with $w=0$ and $\lambda=1$.]{Phase space with $w=0$ and $\lambda=1$. The shaded region represents the part of the phase space where the universe undergoes an accelerated expansion.}
    \label{fig:copxsigmalambda1}
\end{figure}

Different values of the parameters $\lambda$ and $w$ lead to different dynamical behaviours of the system and determine whether Point $B$ exists. Let us assume $\lambda>0$ and choose $w$ such that Point $B$ does not exist. In this case, Points $A_{\pm}$ are both unstable, Point $D$ is a saddle, and Point $C$ acts as the attractor (a stable node) and lies in the region corresponding to accelerated expansion. This behaviour is illustrated, for the case $w=0$ (matter domination), in Fig.~\ref{fig:copxsigmalambda1}. The other interesting case to consider is when all fixed points exist; for example, when $w=0$ and $\lambda=2$. As before, Points $A_{\pm}$ are unstable, while Points $C$ and $D$ are both saddle points. The main difference is that Point $B$, which is a stable node, is now the late-time attractor, although it lies outside the region of accelerated expansion; consequently, it does not correspond to a phase of accelerated expansion.

\section{A new approach to interaction models}
\label{sec:nonconstint2}

\subsection{Model and setup}

We are now in a position to apply this approach to a model with a non-constant interaction term $f(y,\Omega)$, or, equivalently, $f(n,\phi)$. Here $n$ denotes the particle number density. In line with Brown's relativistic fluid formulation~\cite{brown1993}, which is in agreement with the first law of thermodynamics, $p=n \partial\rho/\partial n - \rho$, and  we assume $\rho=\rho(n)$, independent of entropy. This is consistent with the fluid equations which imply constant entropy in cosmology. For a linear equation of state $p=w\rho$, this is equivalent to choosing $\rho=n^{w+1}$. This is shown in \cite{Boehmer:2024rqk}, where we derived the cosmological Einstein field equations
\begin{align}
    3H^2 &= \kappa \Bigl(\rho+\frac12 \dot{\phi}^2+V-6 f H \dot{\phi}\Bigr)\,, \label{eq:fried1}\\
    3H^2+2\dot{H} &= -\kappa\Bigl( p+\frac{1}{2} \dot{\phi}^2- V+2f \ddot{\phi}+2\dot{\phi}^2 \frac{\partial f}{\partial \phi} \Bigr)\,, \label{eq:fried2}
\end{align}
and the modified Klein--Gordon (KG) equation
\begin{equation}
    \ddot{\phi} + 3H\dot{\phi} + \frac{\partial V}{\partial\phi} - 6f(3H^2+\dot{H}) + 18nH^2 \frac{\partial f}{\partial n}=0 \,.
    \label{eq:KG}
\end{equation}
Taking the derivative of Eq.~(\ref{eq:fried1}) with respect to time, and solving Eqs.~(\ref{eq:fried2})~and~(\ref{eq:KG}) for $\dot{H}$ and $\ddot{\phi}$, together with solving Eq.~(\ref{eq:fried1}) for $f$, shows that Eqs.~(\ref{eq:fried1})--(\ref{eq:KG}) imply the fluid energy--momentum conservation equation $\dot{\rho}+3H(\rho+p)=0$. This is a highly non-trivial result which follows from the fact that the entire model is based on a consistent variational approach. We also note that the only dependence on the scale factor, $a(t)$, in the field equations arises through the Hubble function and its derivative and that the equations contain both first and second derivatives of the scalar field $\phi$. The effects of the coupling function $f(n,\phi)$ on the cosmological field equations can be interpreted as an additional contribution to the energy-momentum tensor that preserves the conservation equation.

Following~\cite{copeland1998}, we assume the potential to be exponential $V(\phi)=V_0\exp\left(-\kappa\lambda\phi\right)$, where $V_0>0$ is a constant, and $\lambda \geq 0$ is a dimensionless parameter. Since this is an invertible function, it allows the autonomous system of equations to be closed without introducing an additional dynamical variable. We choose the interaction function to be
\begin{align}
    f(n,\phi)=\frac{k}{2\sqrt{6\kappa}}n^{\alpha(1+w)/2}
    V^{-\alpha/2} \,,
    \label{eq:intterm}
\end{align}
where $\alpha$ is a fixed power and $k$ is the coupling constant. This yields the Friedmann constraint
\begin{equation}
    1-x^2-y^2-\sigma^2+k\,x \frac{\sigma^\alpha}{y^{\alpha}}=0 \,.
    \label{fried-non1}
\end{equation}
When setting $\alpha=-2$, we arrive at
\begin{align}
    1-x^2-y^2-\sigma^2+k\,x \frac{y^2}{\sigma^{2}}=0 
    \quad\implies\quad 
    y^2=\frac{\sigma^2\left(1-x^2-\sigma^2\right)}{\sigma^2-kx}\geq 0\,,
    \label{eq:friedmannconstraintnonconst2}
\end{align}
which indicates one can eliminate $y$ from the equations to keep the dynamical system two-dimensional. One finds that the physical phase space bounded by a line, a circle, and a parabola, consisting of a bounded and an unbounded region given by
\begin{align}
    1-x^2-\sigma^2\geq 0\quad&\text{and}\quad \sigma^2-kx>0\,,\\
   \text{or}\quad 1-x^2-\sigma^2\leq 0\quad&\text{and}\quad \sigma^2-kx<0\,.
\end{align}
The regions intersect a single point $(x,\sigma)$ with coordinates
\begin{alignat}{2}
    &\left(\frac{1}{2} \left(\sqrt{k^2+4}-k\right),\sqrt{\frac{k}{2} (\sqrt{k^2+4}-k)}\right)
    &\quad \text{when} &\quad k>0\,,
    \label{eq:intcoordkplus2}\\
    &\left(-\frac{1}{2} \left(\sqrt{k^2+4}+k\right),\sqrt{-\frac{k}{2} \left(\sqrt{k^2+4}+k\right)}\right)
    &\quad \text{when} &\quad k<0\,.
    \label{eq:intcoordkminus2}
\end{alignat}
Trajectories may traverse this point to pass from one part of the region to the other. The dynamical system obtained for arbitrary $w$ is rather convoluted and difficult to analyse in full generality. However, the system has the desirable property that, in the limit $k\to 0$, the model reduces to the one studied in~\cite{copeland1998}, but expressed in $(x,\sigma)$ coordinates, as discussed in Section~\ref{sec:copelandxsigma}.

To make concrete qualitative statements about this model, let us fix $w$ to zero, that is, in the matter-dominated universe. Under this assumption, the dynamical equation for the variable $x$ is
\begin{equation}
    x'=\frac{2 \sigma ^4 \left(x^2-1\right) \left(6 x-\sqrt{6} \lambda \right)+2 \sigma ^6 \left(3 x-\sqrt{6} \lambda \right)+k\mathcal{A}+k^2\mathcal{B}}{4 \sigma ^4-8 k \sigma ^2 x+k^2 \left(\left(\sigma ^2-1\right)^2+x^4+2 \left(\sigma ^2+1\right) x^2\right)}, \label{eq:xeqna}
\end{equation}
where $\mathcal{A}$ and $\mathcal{B}$ are given by
\begin{align}
    \mathcal{A}=&\,\sigma^2\left(\left(x^2-1\right) \left(x \left(\sqrt{6} \lambda  \left(3 x^2+1\right)-6
   x\right)-12\right)\right.
   \nonumber\\
   & \left.+\sigma ^2 \left(x^2 \left(4 \sqrt{6} \lambda  x+9\right)-15\right)+\sigma ^4 \left(\sqrt{6} \lambda  x+3\right)\right),\\
   \mathcal{B}=&\,x \left(6 \sigma ^4-3 \sigma ^2-\sqrt{6} \lambda  x^5+3 x^4+3 \sigma ^2 x^2+\sqrt{6} \lambda  \left(\sigma ^2-1\right)^2 x-3\right);
\end{align}
while the dynamical equation for the variable $\sigma$ is
\begin{equation}
    \sigma'=\frac{12 \sigma ^5 \left(\sigma ^2+2 x^2-1\right)+k\mathcal{C}+k^2\mathcal{D}}{2\left(4 \sigma ^4-8 k \sigma ^2 x+k^2 \left(\left(\sigma ^2-1\right)^2+x^4+2 \left(\sigma ^2+1\right) x^2\right)\right)} \,,
    \label{eq:sigmaeqna}
\end{equation}
where the terms $\mathcal{C}$ and $\mathcal{D}$ are
\begin{align}
    \mathcal{C}&=2 \sigma ^3 \left(\sqrt{6} \lambda  \left(\sigma ^2-1\right)^2+3 \sqrt{6} \lambda  x^4-12 x^3+4 \sqrt{6} \lambda  \left(\sigma ^2-1\right) x^2\right)\,,\\
    \mathcal{D}&=\sigma  \left(9 \left(\sigma ^2-1\right)^2-4 \sqrt{6} \lambda  x^5+9 x^4
    -4 \sqrt{6} \lambda  \left(\sigma ^2-1\right) x^3+6 \left(\sigma ^2-1\right) x^2\right),
\end{align}
respectively. The acceleration equation is given by
\begin{equation}
    \frac{\dot{H}}{H^2}=\frac{-6 \left(\sigma ^6+2 \sigma ^4 x^2\right)+k\mathcal{E}+k^2\mathcal{F}}{4 \sigma ^4-8 k \sigma ^2 x+k^2 \left(\left(\sigma ^2-1\right)^2+x^4+2 \left(\sigma ^2+1\right) x^2\right)}\,,
\end{equation}
with two more terms $\mathcal{E}$ and $\mathcal{F}$
\begin{align}
    \mathcal{E}&=-\frac{\mathcal{C}}{2\sigma}+12 x\sigma^2\,,\\
   \mathcal{F}&=2 \left(-3 \sigma ^4+\sigma ^2 \left(x^2 \left(\sqrt{6} \lambda  x-3\right)+6\right)+x^3 \left(\sqrt{6} \lambda 
   \left(x^2-1\right)-3 x\right)-3\right) \,.
\end{align}
Finally, the effective equation-of-state parameter $\widetilde{w}$ is given by
\begin{equation}
   \widetilde{w}=\frac{k^2 (\mathcal{D}/\sigma)+ k (\mathcal{C}/\sigma)+12 \sigma ^4
   \left(\sigma ^2+2 x^2-1\right)}{12 \sigma ^4-24 k \sigma ^2 x+3k^2 \left( \sigma ^4+2 \sigma ^2 \left(x^2-1\right)+ \left(x^2+1\right)^2\right)}\,,
\end{equation}
which as $k\to 0$ matches Eq.~\eqref{eq:weffcopxsigma}.

\subsection{Fixed points and stability}

To determine the fixed points of the system, we follow the standard procedure of finding solutions to $x'=0$ and $\sigma'=0$. This leads to the critical points listed in Table~\ref{tab:fixedpointsnonconstintxsigma}, with their stability properties given in Table~\ref{tab:eigenvaluesnonconstint2}, which exhibit the following features. We first note that Point $O$ does not appear in the analysis presented in Section~\ref{sec:copelandxsigma}. The value of the deceleration parameter, $q$, at this point corresponds to a universe undergoing very rapid decelerating expansion. As will be shown below, if one imposes $\widetilde{w}\in(-1,1]$, this point always lies outside the physically relevant phase space. 

\begin{table}[!b]
    \centering
    {\begin{tabular}{@{}lccccc@{}}\hline
             Point & Coordinates $(x,\sigma)$ & Existence & $\widetilde{w}$ & $q$ \\\hline 
            Point $O$ & $(0,0)$ & for all $\lambda$, $k$ & $3$ & $5$\\
            Point $A_{-}$ & $(-1,0)$ & for all $\lambda$, $k$ & $1$ & $2$ 
            \\
            Point $A_{+}$ & $(1,0)$ & for all $\lambda$, $k$ & $1$ & $2$ \\
            Point $B$ & $\displaystyle\left(\sqrt{\frac{3}{2}}\frac{1}{\lambda },0\right)$ & \makecell[l]{
             for $k>0$, $0< \lambda\leq \sqrt{3/2}$  \\[1ex] 
            for $k<0$, $\lambda\geq \sqrt{3/2}$
             } & $\displaystyle 3-\frac{12}{2 \lambda ^2+3}$ & $\displaystyle 5-\frac{18}{2 \lambda ^2+3}$ \\[2ex]
            Point $C_{+}$ & $\displaystyle\left(\sqrt{\frac{3}{2}}\frac{1}{\lambda },\widehat{\sigma}_+\right)$ & see Fig.~\ref{fig:regionplotxsigma_alphaminus2} & $0$ & $1/2$\\[2ex]
            Point $C_{-}$ & $\displaystyle\left(\sqrt{\frac{3}{2}}\frac{1}{\lambda },\widehat{\sigma}_-\right)$ & see Fig.~\ref{fig:regionplotxsigma_alphaminus2} & $0$ & $1/2$\\[2ex]
            Point $D$ & $\left(0,1\right)$ & for all $\lambda$, $k$ & $0$ & $1/2$\\
            Point $E$ & $\displaystyle \left(-\frac{3}{2}\sqrt{\frac{3}{2}}\frac{1}{\lambda},\sqrt{-\frac{3}{2}\sqrt{\frac{3}{2}}\frac{k}{\lambda}}\right)$ & for $k<0$, $\lambda>0$ & $0$ & $1/2$
             \\\hline
        \end{tabular}}
    
\caption{Critical points for Eqs.~\eqref{eq:xeqna}~and~\eqref{eq:sigmaeqna}. The existence conditions indicate the conditions for each critical point to exist within the physical phase space.\\ Here \(\displaystyle \widehat{\sigma}_{\pm}=\frac{\sqrt{4 \lambda ^2-3 \sqrt{6} k \lambda -12\pm\sqrt{54 k^2 \lambda ^2+16 \left(\lambda ^2-3\right)^2+24 \sqrt{6} k \left(\lambda ^2+2\right) \lambda }}}{2
   \sqrt{2} \lambda }.\)}
\label{tab:fixedpointsnonconstintxsigma}
\end{table}

Points $A_{\pm}$ correspond to solutions dominated by the kinetic energy of the scalar field and exist for all values of $\lambda$ and $k$. At these points, $\widetilde{w}=1$, implying that the universe behaves as stiff matter and that no accelerated expansion occurs. For Point $B$ to correspond to a solution exhibiting accelerated expansion, one requires $q<0$, which implies $0<\lambda <\sqrt{3/10}$. Moreover, the value of the effective equation-of-state parameter $\widetilde{w}$ at this point remains within the interval $[-1,1]$ only when $0<\lambda \leq \sqrt{3}/2$. Point $C_{+}$ exists when
\begin{equation}
    \frac{1}{\sqrt{2}}<\lambda \leq \sqrt{3}\;\;\text{and}\;\; k>0\,,\;\;\; 
    \text{or}\;\;\;  \lambda >\sqrt{3}\;\;\text{and}\;\; 2
    \sqrt{6} \lambda ^2-2 \sqrt{30} \sqrt{2 \lambda ^2-1}+9 \lambda k+4 \sqrt{6}>0\,.
\end{equation}
It represents a scaling solution, since the effective equation of state coincides with the matter equation of state, $\widetilde{w}=w=0$. As indicated by the value of the deceleration parameter, accelerated expansion cannot occur at this point. We also note that when $\lambda=1/\sqrt{2}$, Point $C_{+}$ degenerates into Point $B$ because $\widehat{\sigma}_+$ vanishes. Point $C_{-}$ exists when
\begin{equation}
    \lambda >\sqrt{3}
    \quad\text{and}\quad
    \frac{2 \sqrt{60 \lambda ^2-30}}{9 \lambda } -
    \frac{2 \left(\sqrt{6} \lambda ^2+2\sqrt{6}\right)}{9 \lambda }
    \leq k<0\,,
\end{equation}
and is always a saddle point. Point $D$ corresponds to a matter dominated universe with $\sigma^2=1$ and exists as a critical point for all values of $\lambda$ and $k$. It is always a saddle and, as noted previously in Section~\ref{sec:copelandxsigma}, behaves analogously to Point $O$ in $(x,y)$ coordinates. At this point the scalar field is absent, since both its kinetic and potential energy contributions vanish. Moreover, this point lies at the intersection of the two regions of the phase space for $k=(2\lambda^2-3)/(\lambda\sqrt{6})$, or, equivalently, $\lambda =\sqrt{3/8} (k\pm \sqrt{k^2+4})$. Point $E$ exists only when $k<0$ and $\lambda>0$, and is a stable node.

\begin{table}[!h]
    \centering
    {\begin{tabular}{@{}lcl@{}}\hline
             Point & Eigenvalues & Classification  \\ \hline 
              Point $O$ & $9/2\,,\quad -3$& 
             saddle \\
             Point $A_{-}$ & $3/2\,,\quad 3+\sqrt{6}\lambda$& 
             unstable node\\ 
            Point $A_{+}$ & $3/2\,,\quad 3-\sqrt{6}\lambda$& 
             \makecell[l]{
             unstable if $0<\lambda <\sqrt{3/2}$ \\
             saddle if $\lambda >\sqrt{3/2}$
             }\\[3ex]  
             Point $B$ & 
             $3-\frac{18}{2 \lambda ^2+3}\,,\quad \frac{9}{2}-\frac{18}{2 \lambda ^2+3}$ & 
             \makecell[l]{
             stable node if $0<\lambda <1/{\sqrt{2}}$\\
             saddle if $1/\sqrt{2}<\lambda <\sqrt{3/2}$\\
             unstable node if $\lambda >\sqrt{3/2}$ 
             }\\[4ex]  
             Point $D$ &$3\,,\quad -3/2$& 
             saddle\\
             Point $E$ &$-15/2\,,\quad -9$& 
             stable node
             \\\hline
        \end{tabular}}
    \caption{Stability of the critical points for system given in Eqs.~\eqref{eq:xeqna}~and~\eqref{eq:sigmaeqna}. Here $\lambda$ is assumed to be nonnegative. For properties of Point $C$, see Fig.~\ref{fig:regionplotxsigma_alphaminus2}.}
    \label{tab:eigenvaluesnonconstint2}
\vspace*{-4pt}
\end{table}

\begin{figure}[h!]
    \centering
    \includegraphics[width=0.98\textwidth]{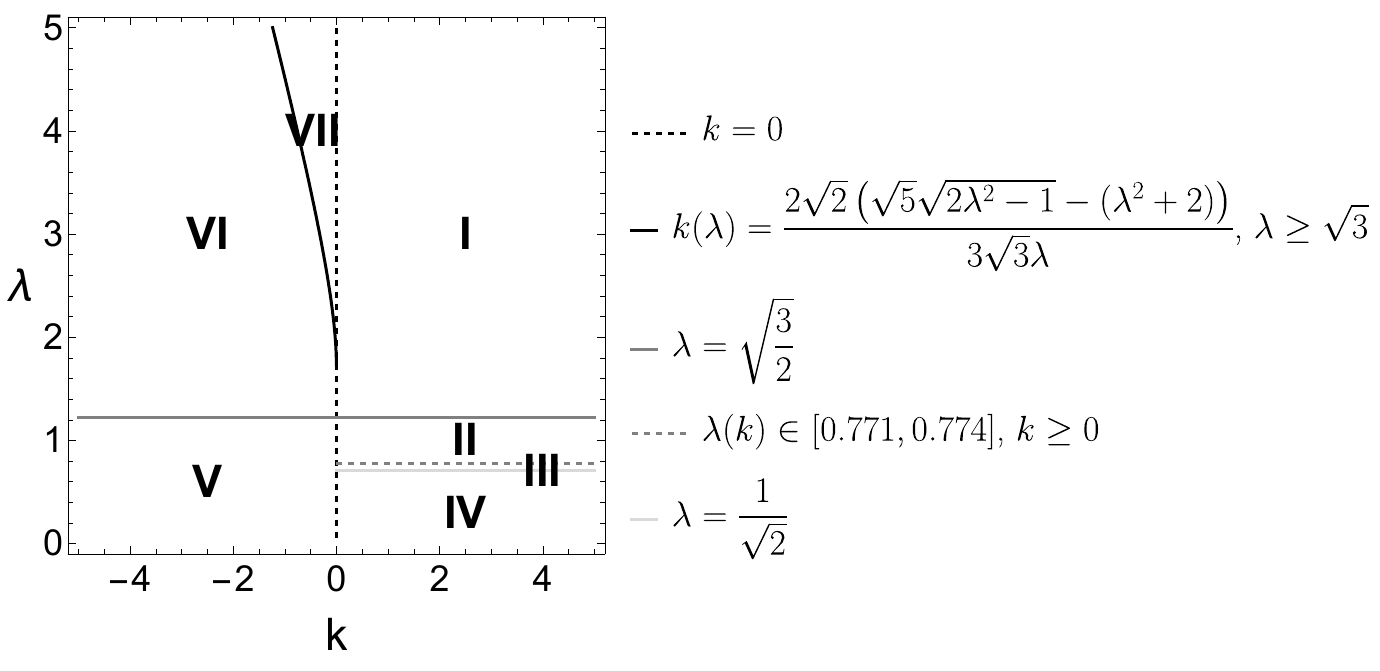}
    \caption[Existence and stability regions in $(k,\lambda)$-plane for $w=0$.]{Existence and stability regions in $(k,\lambda)$-plane for $w=0$. The plotted curves follow from the stability criteria given in Table~\ref{tab:eigenvaluesnonconstint2}. Note that: in Region III, Point $C_+$ exists and is a stable node; in Region I and II, Point $C_+$ exists and is a stable spiral; Point $C_{-}$ exists only in Region VII and is a saddle point, whereas, in the same region, Point $C_{+}$ also exists and is a stable spiral. The boundary between Region II and Region III is not a straight line, but a non-constant function $\lambda(k)$ for $k>0$, which can be characterised as the root of a certain polynomial with integer coefficients of degree $68$ in $\lambda$ and $32$ in $k$. As $k\to 0^{+}$ one finds $\lambda \to \frac{1}{22}\sqrt{3(259-5\sqrt{1057})}\approx 0.773164$, whereas as $k\to \infty$ one obtains $\lambda \to 5/\sqrt{42}\approx 0.771517$.}
    \label{fig:regionplotxsigma_alphaminus2}
\end{figure}

Before analysing the critical points, we remark that, similarly to the case $\alpha=2$, discussed in~\cite{Boehmer:2024rqk}, the point in the physical phase space connecting the two main regions (see Eqs.~\eqref{eq:intcoordkplus2} and \eqref{eq:intcoordkminus2}) corresponds to an interesting point of the dynamical system. It is not possible to assign a unique value of $q$ or $\widetilde{w}$ at this point, since neither quantity has a well-defined limit as $x$ and $\sigma$ approach the intersection. Nevertheless, along any given trajectory passing through this point, the limits of $q$ and $\widetilde{w}$ do exist. In particular, a trajectory in the physical phase space satisfying $-1<\widetilde{w}\le1$ and traversing the intersection will also satisfy $-1<\widetilde{w}\le1$ at that point.

\subsection{Phase space diagrams and physical interpretation.} 

For varying values of $\lambda$ and $k$, we obtain different models, as shown in Fig.~\ref{fig:regionplotxsigma_alphaminus2}. Here we present some of the different cases and illustrate their cosmological meaning. We omit details for Regions V, VI, and VII (corresponding to $k<0$) since they are of no physical interest.

In Region I, there are five fixed points within the physical phase space, as determined by the Friedmann constraint. However, the values of the effective equation-of-state parameter $\widetilde{w}$ are not physical at all fixed points (for example, at Point $O$ one has $\widetilde{w}=3$). Consequently, an additional restriction on $\widetilde{w}$ is required to identify the physically relevant phase space. This restricts the admissible trajectories. A trajectory originating near $A_{-}$ reaches the saddle point $D$, traverses the phase space, and is eventually attracted to Point $C_{+}$. For a trajectory to approach Point $A_{+}$, it would need to pass through a region where $\widetilde{w}<-1$. We note that Point $C_{+}$ again corresponds to a scaling solution, although it does not describe a phase of accelerated expansion. Since this case contains only a single attractor (stable spiral), the dynamical evolution is unique. For a solution with $\ddot{a}>0$, we need the values of $\lambda$ and $k$ to lie within Region IV. The restricted physical phase space permits only a small number of trajectories. In particular, two types of evolution arise. A trajectory starting near Point $A_{-}$ first approaches the saddle Point $D$, then crosses the phase space and eventually converges to Point $B$, which lies in the region of accelerated expansion. Alternatively, there is a simpler evolution in which the trajectory begins in the vicinity of Point $A_{+}$ and also terminates at Point $B$, but this is not physically interesting. We note that physically admissible trajectories producing phantom dark energy can also occur within this model.

\begin{figure}[!t]
    \centering
    \includegraphics[width=0.98\textwidth]{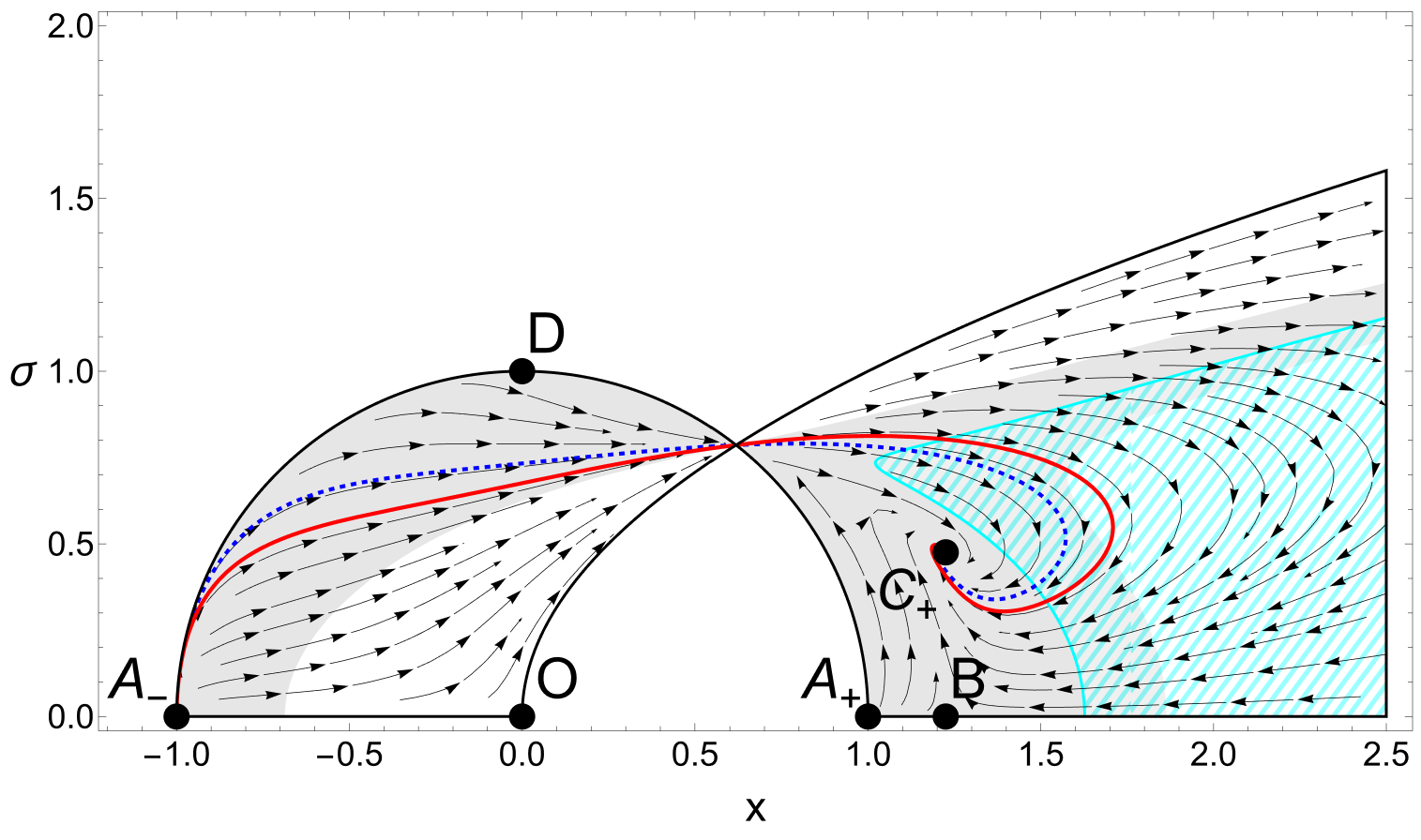}
    \caption[Phase space with $\lambda=1$ and $k=1$.]{The parameter values are $\lambda=1$ and $k=1$. Point $B$ is a saddle and Point $C$ is a stable spiral. The shaded grey region represents the part of the physical phase space where $-1<\widetilde{w}\leq 1$. The hatched cyan region represents the part where universe is accelerating. The dashed blue (initial condition $x(0)=-0.99$ and $\sigma(0)=0.1399$) and solid red (initial condition $x(0)=-0.99$ and $\sigma(0)=0.033$) curves represent two numerical trajectories.}
    \label{fig:lambda1k1_negalpha}
\end{figure}

Region II mainly differs from Region III because of Point $C_{+}$, which is a stable spiral in Region II and a stable node in Region III. As an illustrative example for Region II, let us consider Fig.~\ref{fig:lambda1k1_negalpha}, where $\lambda=k=1$. 
A physically admissible trajectory begins near Point $A_{-}$ and approaches the saddle Point $D$, but never comes close to Point $O$. The trajectory then crosses to the right-hand side of the phase space, entering the region of accelerated expansion, and spirals towards Point $C$. In Fig.~\ref{fig:lambda1k1_negalpha}, two numerical trajectories are shown. The dashed blue trajectory remains entirely within the region $-1<\widetilde{w}\leq 1$. By contrast, the solid red trajectory, once in the region of accelerated expansion, leaves the interval $-1<\widetilde{w}\leq 1$ and enters a phantom regime with $\widetilde{w}<-1$ (in this case $\widetilde{w}\approx -1.17$). 

Let us emphasise that the phantom regime observed is effective in nature. The underlying scalar field retains the canonical kinetic term, and hence no ghost degree of freedom is introduced at the level of the action. The apparent violation of the null energy condition arises from the interaction term which modifies the effective energy–momentum tensor. As a result, the total, effective fluid can display phantom behaviour. The phantom phase is transient, occurring along trajectories that pass through the accelerating region of the phase space before approaching a late-time attractor. The duration and presence of this regime depend on the parameters $\lambda$, $k$, and the chosen initial conditions. This is consistent with other interacting dark energy models, where effective phantom crossing can occur without pathologies in the fundamental degrees of freedom.

\begin{figure}[!tb]
    \centering
    \includegraphics[width=0.8\textwidth]{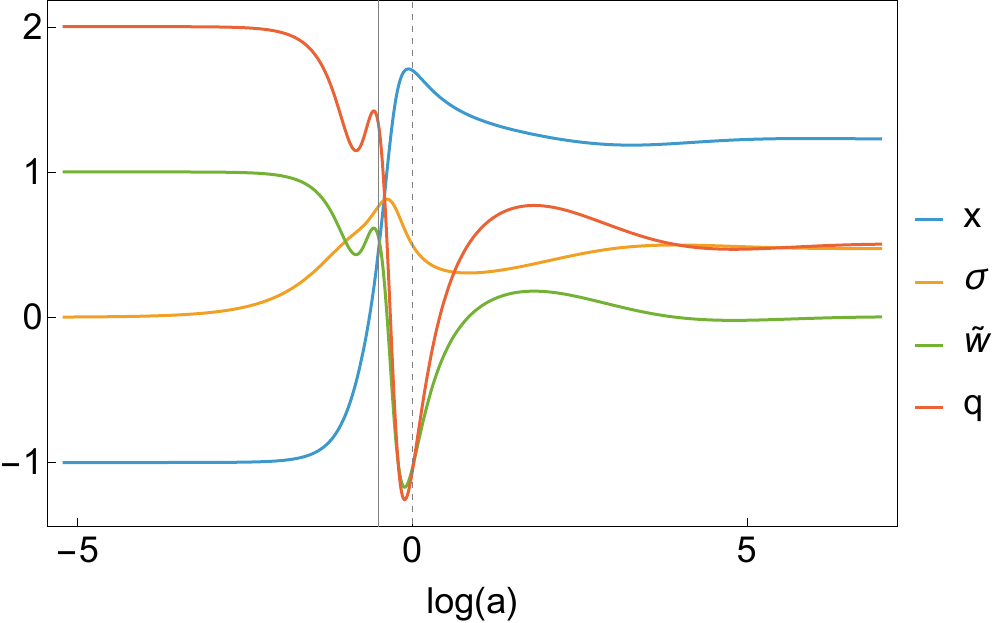}
    \caption[Evolution plot for a trajectory with $\lambda=1$ and $k=1$.]{Evolution of the solutions and physical parameters. The parameter values are $\lambda=1$ and $k=1$. The initial conditions are $x(0)=-0.99$ and $\sigma(0)=0.033$), see solid red trajectory in Fig.~\ref{fig:lambda1k1_negalpha}. The solid grey line represents when the trajectory traverses the intersection between the two regions of the physical phase space; the dashed grey line represents the present time.}
    \label{fig:evolplot}
\end{figure}

The evolution of $x(t)$ and $\sigma(t)$, together with the physical parameters $\widetilde{w}$ and $q$, for this trajectory are displayed in Fig.~\ref{fig:evolplot}. 
This solution exhibits both an early-time inflationary phase and late-time accelerated expansion. The effective equation-of-state parameter crosses the phantom divide at the present epoch (when $\log{a}=0$), while at late times the trajectory approaches Point $C_{+}$, which is matter dominated, as also indicated by the equation-of-state parameter tending to zero. As expected, the system spends the longest time in the vicinity of the fixed points. On the other hand, a trajectory originating near Point $A_{+}$ moves towards the  saddle Point $B$ and subsequently spirals into Point $C$, after a short phase of accelerated expansion. Point $C$ again represents a scaling solution.

\section{Conclusion}

While the cosmological field equations are fixed by a given theory of gravity, the dynamical system derived from them is sensitive to the choice of physical variables used to analyse the system. This choice affects the structure and geometry of the physical phase space and, as we have shown in some cases, can determine whether the system can be easily investigated.

We first reviewed the standard formulation of cosmological models using dimensionless density parameters and then discussed the widely used variables introduced in~\cite{copeland1998} for quintessence models. Solving the Friedmann constraint for the variable $y$ allowed us to reinterpret the well-known exponential potential model in the $(x,\sigma)$ coordinates, providing a useful baseline system for later work. The main motivation for this reformulation arises when considering interacting models with non-trivial coupling terms. For the interaction model studied here, the usual $(x,y)$ variables remain suitable when $\alpha=2$, as discussed in detail in~\cite{Boehmer:2024rqk}. However, for the case $\alpha=-2$ or other values, this choice of variables is not suitable. Solving the Friedmann constraint for $y$ instead of $\sigma$ leads to a more convenient form of the dynamical equations. Our phase space analysis shows that the resulting models exhibit a rich mathematical structure depending on the parameter values of $k$ and $\lambda$. An interesting feature of this analysis is that the physically relevant phase space is determined not only by the Friedmann constraint, but also by imposing constraints on the physically admissible values of the effective equation-of-state parameter. 

The examples considered allow us to work in a two-dimensional phase space. However, in other cases, such as those discussed in~\cite{Ashi:2025dba}, the introduction of another variable $z$, depending on the Hubble function $H$, is necessary since terms containing the Hubble function in the Friedmann constraint cannot be eliminated; this variable was first used in~\cite{boehmer2008} and later employed in~\cite{boehmer2010}. It is also worth noting that we have assumed the potential to be of the exponential form throughout. A natural extension of our work would explore other potentials, like the power-law potential, which would in turn increase the dimensionality of the phase space. 

Our results highlight that an appropriate choice of variables can reveal the underlying structure of the model and enable the study of interaction models that would otherwise be intractable. This perspective is particularly relevant for more complicated interacting dark energy models and modified gravity theories.


\addcontentsline{toc}{section}{References}
\begin{thebibliography}{99}

\bibitem{Wiggins2003}
Wiggins S. 2003.
\textit{Introduction to Applied Nonlinear Dynamical Systems and Chaos},
2nd ed., Texts in Applied Mathematics, vol.~2,
Springer, New York.

\bibitem{Strogatz2024} Strogatz SH. 2024. \textit{Nonlinear dynamics and chaos: with applications to physics, biology, chemistry, and engineering}. 
3rd ed. Boca Raton, FL: CRC Press. See \href{https://doi.org/10.1201/9780429398490}{https://doi.org/10.1201/9780429398490}.

\bibitem{wainwright1997}
Wainwright J, Ellis GFR (eds.). 1997.
\textit{Dynamical Systems in Cosmology}.
Cambridge, UK: Cambridge University Press.

\bibitem{coley2003}
Coley AA. 2003.
\textit{Dynamical Systems and Cosmology}.
Dordrecht: Springer.

\bibitem{leon2011}
Leon G, Fadragas CR. 2011.
\textit{Cosmological Dynamical Systems}.
Saarbrücken: LAP Lambert Academic Publishing. 

\bibitem{bahamonde2018}
Bahamonde A, Böhmer CG, Carloni S, Copeland EJ, Fang W, Tamanini N. 2018.
\textit{Dynamical Systems Applied to Cosmology: Dark Energy and Modified Gravity}.
Cham: Springer. See
\href{https://doi.org/10.1016/j.physrep.2018.09.001}{https://doi.org/10.1016/j.physrep.2018.09.001}.

\bibitem{PlanckCollabVI} Planck Collaboration VI. 2020 Planck 2018 results. VI. Cosmological parameters. 
\textit{Astron. Astrophys.} \textbf{641}, A6. See \href{https://doi.org/10.1051/0004-6361/201833910}{https://doi.org/10.1051/0004-6361/201833910}.

\bibitem{efstathiou2020} Efstathiou G, Gratton S. 2020. The evidence for a spatially flat Universe. \textit{Mon. Not. R. Astron. Soc.} \textbf{496}, 1, pages L91–L95. See \href{https://doi.org/10.1093/mnrasl/slaa093}{https://doi.org/10.1093/mnrasl/slaa093}.

\bibitem{Boehmer2017a}
Böhmer CG. 2017.
\textit{Introduction to General Relativity and Cosmology},
World Scientific Publishing, Singapore.

\bibitem{Boehmer2017} Böhmer CG, Chan N. 2017. Dynamical systems in cosmology. In 
\textit{Dynamical and Complex Systems} (eds S Bullett, T Fearn, F Vivaldi), ch. 4, pp. 121--156. London, UK: World Scientific. See 
\href{https://doi.org/10.1142/9781786341044_0004}{https://doi.org/10.1142/9781786341044\_0004}.

\bibitem{Perez2014} Perez J, F\"{u}zfa A, Carletti T, Heghran L, Petit G. 2014. The Jungle Universe: coupled cosmological models in a Lotka–Volterra framework. \textit{Gen. Relativ. Gravit.} \textbf{46}, 1753. See \href{https://doi.org/10.1007/s10714-014-1753-8}{https://doi.org/10.1007/s10714-014-1753-8}.

\bibitem{amendola2000}
Amendola L. 2000.
Coupled quintessence.
\textit{Phys Rev D}. \textbf{62}, 043511. See
\href{https://doi.org/10.1103/PhysRevD.62.043511}{https://doi.org/10.1103/PhysRevD.62.043511}.

\bibitem{copeland2006}
Copeland EJ, Sami M, Tsujikawa S. 2006.
Dynamics of dark energy.
\textit{Int. J. Mod. Phys. D}. \textbf{15}, 11. See
\href{https://doi.org/10.1142/S021827180600942X}{https://doi.org/10.1142/S021827180600942X}.

\bibitem{copeland1998}
Copeland EJ, Liddle AR, Wands D. 1998.
Exponential potentials and cosmological scaling solutions.
\textit{Phys Rev D}. \text{57}, 4686. See
\href{https://doi.org/10.1103/PhysRevD.57.4686}{https://doi.org/10.1103/PhysRevD.57.4686}.

\bibitem{Boehmer:2024rqk}
Böhmer CG, d'Alfonso del Sordo A. 2024.
Cosmological fluids with boundary term couplings.
\textit{Gen. Relativ. Gravit.} \textbf{56}, 75. See \href{https://doi.org/10.1007/s10714-024-03260-6}{https://doi.org/10.1007/s10714-024-03260-6}.

\bibitem{Ashi:2025dba}
Ashi HA, Böhmer CG, d'Alfonso del Sordo A, Jensko E. 2025.
Cosmological dynamical systems of non-minimally coupled fluids and scalar fields.
\textit{Gen. Relativ. Gravit.} \textbf{57}, 167. See \href{https://doi.org/10.1007/s10714-025-03502-1}{https://doi.org/10.1007/s10714-025-03502-1}.

\bibitem{dAlfonso2025}
d'Alfonso del Sordo A. 2025.
\textit{Special Functions and Dynamical Systems in Cosmology}.
Ph.D. thesis, University College London. See \href{https://discovery.ucl.ac.uk/id/eprint/10214939/}{https://discovery.ucl.ac.uk/id/eprint/10214939/}.

\bibitem{brown1993} Brown JD. 1993. Action functionals for relativistic perfect fluids. \textit{Class. Quantum Grav.} \textbf{10}, 1579. See \href{https://doi.org/10.1088/0264-9381/10/8/017}{https://doi.org/10.1088/0264-9381/10/8/017}.

\bibitem{boehmer2008}
Böhmer CG, Caldera-Cabral G, Lazkoz R, Maartens R. 2008. 
Dynamics of dark energy with a coupling to dark matter.
\textit{Phys. Rev. D}. \textbf{78}, 023505. See
\href{https://doi.org/10.1103/PhysRevD.78.023505}{https://doi.org/10.1103/PhysRevD.78.023505}

\bibitem{boehmer2010}
Böhmer CG, Caldera-Cabral G, Lazkoz R, Maartens R. 2010.
Quintessence with quadratic coupling to dark matter.
\textit{Phys. Rev. D}. \textbf{81}, 083003. See
\href{https://doi:10.1103/PhysRevD.81.083003}{https://doi:10.1103/PhysRevD.81.083003}

\end{thebibliography}
\end{document}